\documentclass[sigconf]{aamas}



\usepackage{balance} % for balancing columns on the final page
\usepackage{graphicx}
\usepackage{quotes}
\graphicspath{ {Figures/} }
\usepackage{algorithm}
\usepackage [noend] {algpseudocode}

\usepackage {algxpar}
\usepackage {tabularx}
\usepackage{etoolbox}
\usepackage{hyperref}
\doi{DEZC1973}

\makeatletter\gdef\@copyrightpermission{  \begin{minipage}{0.2\columnwidth}   \href{https://creativecommons.org/licenses/by/4.0/}{\includegraphics[width=0.90\textwidth]{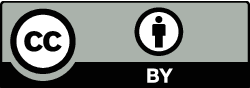}}  \end{minipage}\hfill  \begin{minipage}{0.8\columnwidth}   \href{https://creativecommons.org/licenses/by/4.0/}{This work is licensed under a Creative Commons Attribution International 4.0 License.}  \end{minipage}  \vspace{5pt}}\makeatother

\setcopyright{ifaamas}
\acmConference[AAMAS '26]{Proc.\@ of the 25th International Conference
on Autonomous Agents and Multiagent Systems (AAMAS 2026)}{May 25 -- 29, 2026}
{Paphos, Cyprus}{C.~Amato, L.~Dennis, V.~Mascardi, J.~Thangarajah (eds.)}
\copyrightyear{2026}
\acmYear{2026}
\acmDOI{}
\acmPrice{}
\acmISBN{}

\acmSubmissionID{141}

\title[AAMAS-2026 Formatting Instructions]{Stigmergic Swarming Agents for Fast Subgraph Isomorphism}

\author{\href{https://orcid.org/0000-0002-3434-5088}{H. Van Dyke Parunak}}
\affiliation{
  \institution{ABC Research}
  \city{Ann Arbor, MI}
  \country{USA}}
\email{van.parunak@gmail.com}

\begin{abstract}
Maximum partial subgraph isomorphism compares two graphs (nodes joined by edges) to find a largest common subgraph. A common use case, for graphs with labeled nodes, seeks to find instances of a \textit{query} graph with $q$ nodes in a  (typically larger) \textit{data} graph with $d$ nodes. The problem is NP-complete, and naïve solutions are exponential in $q + d$. The fastest current heuristic has complexity $O(d^2)$. This paper outlines ASSIST (Approximate Swarming Subgraph Isomorphism through Stigmergy), inspired by the ant colony optimization approach to the traveling salesperson. After peering (identifying matching individual nodes in query and data) in time $O(q\cdot log(d))$, the time required for ASSIST's iterative subgraph search, the combinatorially complex part of the problem, is linear in query size and constant in data size. ASSIST can be extended to support matching problems (such as temporally ordered edges, inexact matches, and missing nodes or edges in the data graph) that frustrate other heuristics. 
\end{abstract}



\newcommand{\BibTeX}{\rm B\kern-.05em{\sc i\kern-.025em b}\kern-.08em\TeX}

\algnewcommand\algorithmicswitch{\textbf{switch}}
\algnewcommand\algorithmiccase{\textbf{case}}
\algnewcommand\algorithmicassert{\texttt{assert}}
\algnewcommand\Assert[1]{\State \algorithmicassert(#1)}%
\algdef{SE}[SWITCH]{Switch}{EndSwitch}[1]{\algorithmicswitch\ #1\ \algorithmicdo}{\algorithmicend\ \algorithmicswitch}%
\algdef{SE}[CASE]{Case}{EndCase}[1]{\algorithmiccase\ #1}{\algorithmicend\ \algorithmiccase}%
\algtext*{EndSwitch}%
\algtext*{EndCase}%
\algtext*{EndIf}
\algtext*{EndFor}

\begin{document}

%%% The following commands remove the headers in your paper. For final 
%%% papers, these will be inserted during the pagination process.

\pagestyle{fancy}
\fancyhead{}

%%% The next command prints the information defined in the preamble.

\maketitle 

%%%%%%%%%%%%%%%%%%%%%%%%%%%%%%%%%%%%%%%%%%%%%%%%%%%%%%%%%%%%%%%%%%%%%%%%

\section{Introduction}
The maximum partial subgraph isomorphism problem is: given two graphs $G_1, G_2$, find a largest subgraph of $G_1$ that is isomorphic to a subgraph of $G_2$. The problem is well-defined whether the graph nodes of the graphs are labeled from some set \textbf{L} of size $|\textbf{L}| > 1$  (with some labels possibly repeated), or whether they are unlabeled (equivalently, labeled from a set of size 1). In both cases, the problem is NP-complete. (Our solution requires labeled graphs, but allows some labels to be repeated.) 

For example, chemists frequently seek shared structures in large organic molecules, where nodes are the atoms in a molecule (e.g., C, H, O, ...), and edges are chemical bonds between them. Finding a common structure of six atoms between two organic molecules, each of 50 atoms, with a na\"ive enumeration requires over $10^{17}$ comparisons, motivating chemists to pursue research in subgraph isomorphism algorithms \cite{RN1}.

In addition to molecular design, many other applications would benefit from the ability to compare even larger graphs. For example:
\begin{enumerate}
	\item NOSQL databases maintain networks of relationships among millions of data items. A natural query is a graph of possible relations \cite{RN2}, and analysts seek a largest subgraph in the query that occurs in the data \cite{RN3}.
	
    \item Financial transaction data is a primary data source for detecting financial crime such as money laundering, if its petabytes of data can be searched efficiently for patterns detailing known transactional behaviors \cite{A3ML}.
    
    \item IP packet data is an important resource for network monitoring. A connection is a five-tuple <source-IP, source-port, dest-IP, dest-port, protocol>, and an edge joins one connection C1 to another C2 just in case the start time of C1 is earlier than that of C2 and at least one IP address is repeated between them. Such graphs will have millions of connections for reasonable periods of time. Common subgraphs between graphs representing different networks, or the same network at different times, highlight shared behaviors that might indicate common actors \cite{leie17, godi10}.
    
    \item A narrative space \cite{RN5, RN4} fuses many possible causal trajectories, and enables highly-understandable social simulations. Such graphs, which can contain hundreds of nodes, enable analysts to visualize and interact in both forensic and forecasting problems, but authoring them is time consuming. Many domains maintain collections of narratives about specific past events \cite{RN10, RN9, RN8, RN7, RN6}. Fusing these narratives at their common subgraphs would yield a narrative space for their domain, greatly accelerating the analytic process.
    
    \item Fusing patient records in health care can generate a causal model of diagnoses, treatments, and outcomes for resource forecasting and fraud detection \cite{fare24}.
    
    \item Image recognition makes use of feature graphs that capture adjacency information among different features \cite{jian04, lu21}. More efficient subgraph isomorphism algorithms would allow powerful new image search capabilities.
    
\end{enumerate}
    
In these and similar cases, the graphs range from 10$^3$ to 10$^6$ nodes or even more. With even the 50-node graphs of organic chemistry posing computational limits, a faster algorithm is clearly important.

This paper describes ASSIST (Approximate Swarming Subgraph Isomorphism through STigmergy), a novel heuristic inspired by ant colony optimization (ACO) \cite{RN27}. "Stigmergy" refers to coordination of agents by making and sensing changes in a shared environment \cite{gras59}, rather than by direct inter-agent messaging. It is inspired by social insects such as ants and termites, who coordinate their work by depositing chemicals (“pheromones”) in the environment and making decisions based on the current pheromone strengths in their vicinity. These chemicals evaporate over time, and locations that are reinforced by many ants converge to the selected solution. In nature, these mechanisms enable termites to construct mounds with separate floors and rooms, and ventilation systems to exhaust waste gasses. Ants use them to construct minimal spanning trees joining their nests to food sources. ACO algorithms replace the ants with simple software agents, the pheromones with increments that the agents make to variables on the different locations that they visit, and evaporation with a periodic attenuation of the pheromones by a specified percentage. The results have proven successful in other highly complex problems such as the traveling salesperson. ASSIST applies these techniques to subgraph isomorphism.

In addition to efficient subgraph matching, ASSIST (like other swarming algorithms) demonstrates how stigmergy can integrate partial results obtained by many independent agents without sophisticated coordination mechanisms, inviting its application to other problems of interest to the AAMAS community. 

Section 2 compares ASSIST with other approaches to subgraph isomorphism. Section 3 describes the swarming traveling salesperson (TSP) algorithm that inspires ASSIST. Section 4 describes the ASSIST algorithm. Section 5 reports experimental results.  Section 6 discusses future work. Section 7 concludes.

%\vspace{-3mm}
\section{Related Work}
We compare ASSIST first with other graph matching algorithms, then with other stigmergic systems. Graph matching is an active research area. Convenient surveys include \cite{RN13, RN14, RN12, RN1}. We situate ASSIST in this context along five dimensions, highlighting how its features deliver four key benefits: Tunable Accuracy, Scalability, Robustness, and Accessibility (understandable by non-technical users). Table 1 summarizes the contribution of each feature to the benefits. We discuss the rows in order.

\begin{table}[t]
	\caption{ASSIST compared with other methods for subgraph isomorphism}
	\label{tab:characteristics}
	\begin{tabularx}{\linewidth}{lXX}\toprule
		\textit{ASSIST Feature } & \textit{Competition} & \textit{Benefits of ASSIST} 		\\ \midrule
		Heuristic & Exact \cite{RN15, RN16}  &	Scalability, Robustness\\
		Direct & Transformed \cite{RN17, RN18, RN19, RN20} & 	Robustness, Accessibility\\
		Stochastic & Deterministic \cite{RN15, RN16, RN17, RN18, RN20,  RN24, RN23, RN25, RN26} &	Tunable Accuracy\\
		Multiple Solutions & Single Solution \cite{RN15, RN16, RN17, RN18, RN20,  RN24, RN23, RN25, RN26}  &	Robustness, Accessibility\\
		Incremental & Entire \cite{RN15, RN16, RN17, RN18, RN19, RN20, RN21, RN22} &	Scalability\\  \bottomrule
	\end{tabularx}
\end{table}

ASSIST is a \textit{heuristic}. \textit{Exact methods} (such as Ullman’s pioneering algorithm \cite{RN15}) are guaranteed to find matching subgraphs, but do not scale to large problems, and they are not robust to noise in the query or data. Successive filtering methods \cite{RN16} apply a sequence of exact methods to the data, hoping to reduce its size before attempting to identify subgraphs, but are not robust to incompletely labeled data. 

ASSIST manipulates the query and data \textit{directly}, making it robust to poorly conditioned graphs and accessible to analysts. Spectral methods \cite{RN18, RN17}, Estimation of Distribution Algorithms (EDAs \cite{RN19}), and Optimal Transport approaches \cite{RN20} (with best current complexity O($d^2$)) \textit{transform} the problem (into matrix parameters, probabilistic graphical models, or probability distributions, respectively), which can compromise robustness and accessibility. 

 ASSIST, like EDAs \cite{RN19} and genetic algorithms (GAs) \cite{RN21, RN22}, is \textit{stochastic}, allowing it to escape from local optima, and also permitting tunable accuracy, dynamically trading off probability of detection ($p_d$) and false acceptance rate. Most methods are \textit{deterministic}, giving the same answer each time they run, but are vulnerable to local optima, lacking tunable accuracy.

ASSIST considers \textit{multiple} solutions concurrently, ranking them probabilistically (by the strength of the pheromone field on each one) and thus enhancing accessibility. Most methods produce a \textit{single} matching, which is best by some criterion but may not be robust to corrupted patterns or data. EDAs  and GAs  consider a population of individual competing solutions, allowing alternative matches to emerge, but typically focus down to a single solution, so that analysts never see alternatives. 

ASSIST, like some other methods \cite{RN24, RN23, RN25, RN26}, is \textit{incremental}, starting with local matches within each of the two graphs and expanding the match. Incremental match construction enhances scalability by limiting the effort spent on candidate matches that end up failing. Most methods reason about the \textit{entire} query at once.

Stigmergic swarming, the heart of ASSIST, has been applied successfully in many areas, including the traveling salesperson problem \cite{RN27}, telecommunications routing \cite{RN28}, and geospatial reasoning \cite{RN29}. ASSIST extends these techniques. Unlike geospatial reasoning (but like routing problems) it is not limited to the regular structure of a lattice, but handles arbitrary graphs. Unlike previous graph applications, it has separate pheromone families for nodes and edges, and can be extended to deal with time-sequenced graphs to handle time compactly. 

%\vspace{-3mm}
\section{An Example of Swarming Graph Computation}
One of the most successful applications of swarming agents to graph-theoretic problems is to the traveling salesperson problem (TSP): given a set of points in two dimensions and a road network among them, find the shortest Hamiltonian circuit (visiting every node exactly once and returning to the start). This problem is of great importance in problems such as logistics \cite{RN30} (optimizing fuel usage for truck fleets), telecommunications \cite{RN31}, and wiring plans for printed circuit boards \cite{RN32}, and is successful enough to find widespread commercial use (e.g., www.antoptima.com). 

\begin{figure}[h]
	\centering
	\includegraphics[width=1.0\linewidth]{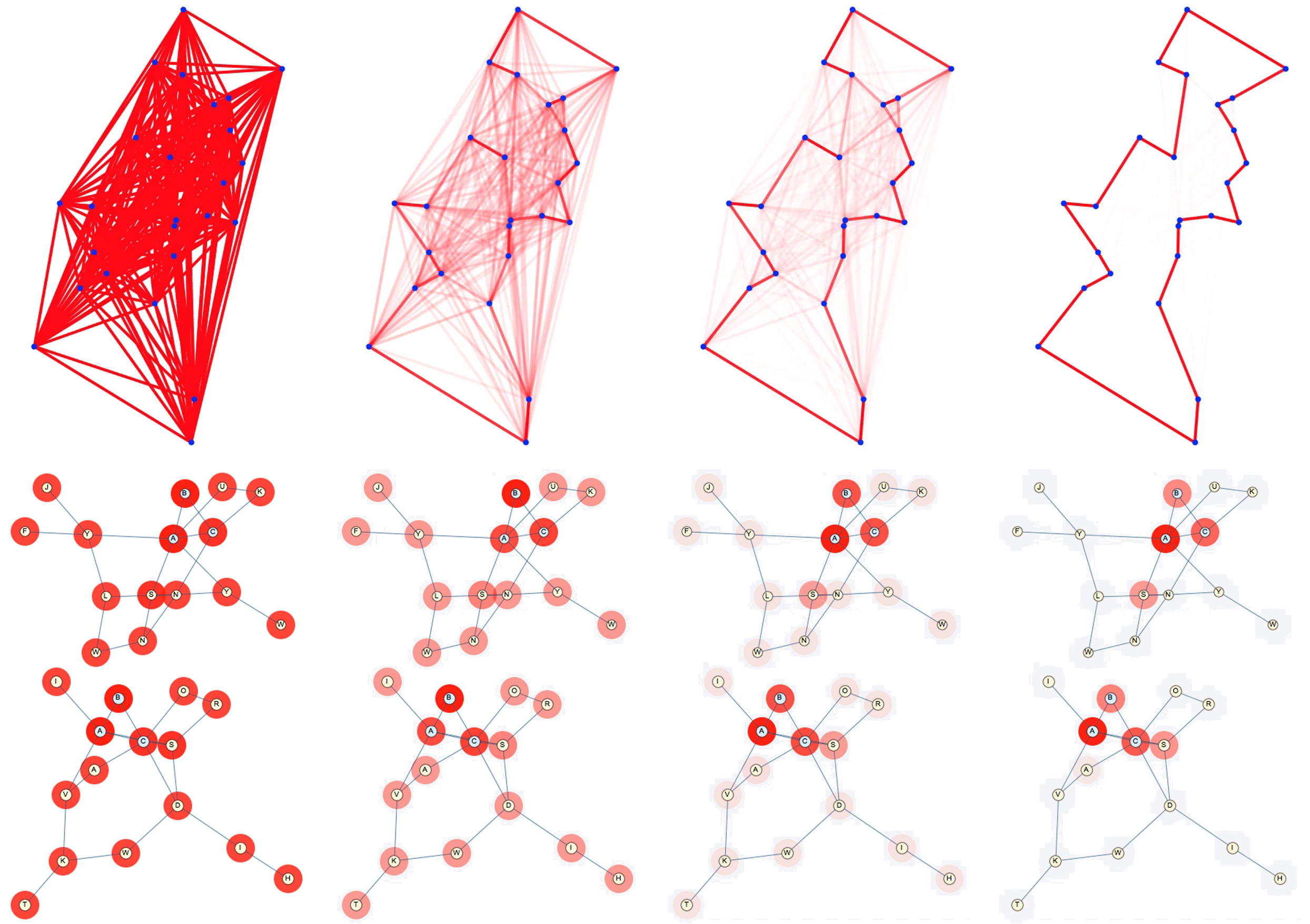}
	\caption{Swarming solutions of the traveling salesperson and subgraph isomorphism}
	\label{fig:aco}
	%\Description{Four stages in the ant colony approach to solving the traveling salesperson problem}
\end{figure}

In these applications, successive waves of software agents (digital ants) explore alternative tours in parallel. Each ant chooses at each step among vertices it has not visited, preferring edges with the strongest pheromones left by previous ants, and returning home when all vertices have been visited. After completing a circuit, each ant deposits pheromone on the edges it has traversed, with strength inversely proportional to the length of its overall path. After each wave of agents explores the graph, all pheromone strengths evaporate, multiplied by $1 - \rho$, $\rho < 1$. Edges that are part of shorter circuits accumulate more pheromone, and attract more agents, while edges that are not visited evaporate, and a highly competitive path emerges (Figure~\ref{fig:aco}, top). Depending on details of the application, the time complexity is O($(n\cdot log(n))/\rho$) \cite{RN33}.

The ASSIST algorithm described in the next section (Figure~\ref{fig:aco}, bottom) uses the same mechanisms of constant pheromone evaporation and selective deposit. Through iteration,  a maximal subgraph emerges (in this case, nodes A-B-C-S). (The edge from $A$ to $S$ in the lower graph is present, but obscured by $C$.)

 \section{The ASSIST Heuristic}
\begin{figure}[h]
	\centering
	\includegraphics[width=0.6\linewidth]{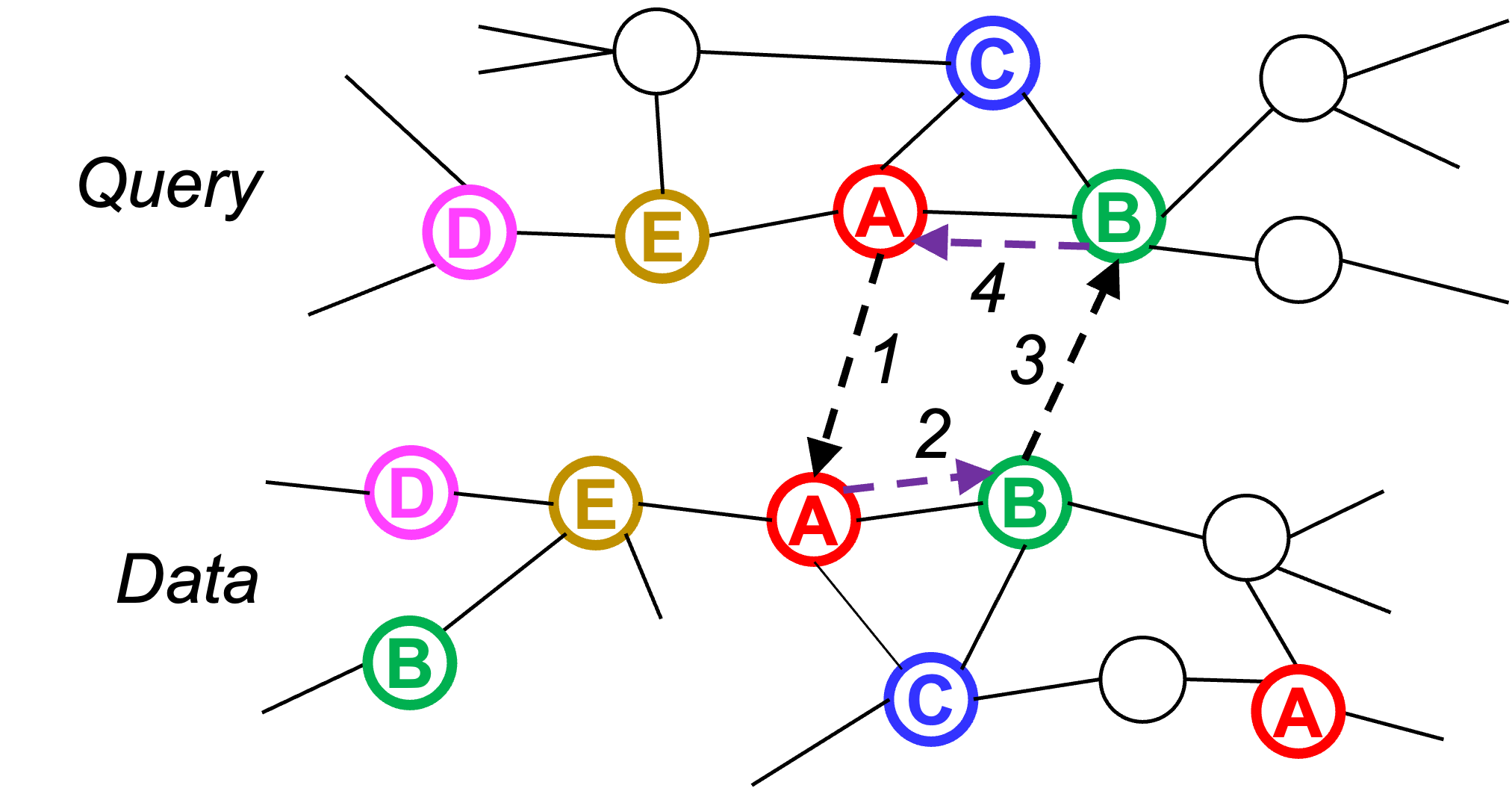}
	\caption{Each agent seeks a matching edge between query and data graphs}
	\label{fig:circuit}
	%\Description{Reinforcement loop (dashed) between common edges in a shared subgraph}
\end{figure}

 Instead of Hamiltonian circuits, ASSIST seek paths between the query and data graphs that traverse matching fragments of both graphs (typically, a single edge). Figure~\ref{fig:circuit} illustrates the movements of a single agent. Starting at a node labeled $A$ in the query, it seeks an $A$ node in the data that has similar neighbors to its original node (1). Then it looks for a neighbor in the data that matches a neighbor of $A$ in the query (B), returns to a $B$ node in the query, and seeks to complete the circuit. If it is successful, it has found a candidate edge in the desired matching, and augments the pheromones on the nodes and edges in both the query and the data. 

Matching edges aggregate into larger subgraphs as swarming agents seek for neighbors of already-matched nodes and reinforce the pheromone levels on those nodes and the edges that join them. Nodes that participate in more than one shared edge are further reinforced. Multiple edges in a shared subgraph reinforce each other, and the larger the subgraph, the more pheromone its nodes and edges accumulate. Pheromones on nodes in either graph that do not have shared edges eventually evaporate to 0. As the pheromones from many agents stabilize, a high quality matching emerges. 

The following sections discuss 1) the data model for the graphs we analyze, 2) the classes and pseudocode of the algorithm, 3) how the graphs are initialized and possible matches identified, 4) the matching of a single edge, and 5) how we detect termination.

\subsection{Data Model}
\label{sect:datamodel1}
For each experiment, we randomly generate three graphs:
\begin{enumerate}%[topsep = 1ex]
	\item A \textit{data} graph represents a graph database that we wish to explore against a query.
	\item A \textit{query} graph represents a set of relations that we are seeking in the data graph. It is typically smaller than the data graph, and may be less specific.
	\item A common \textit{kernel} graph is embedded in both the data graph and the kernel graph. We evaluate ASSIST based on its ability to retrieve the kernel. 
\end{enumerate}
We term such a set of three graphs, a \textit{scenario.}

The kernel is not necessary to the operation of the algorithm, but supports evaluation. Having a known common subgraph in query and data lets us compare the effect of independent variables such as query, data, and vocabulary size or degree of ablation. 

Our experiments use Barab\'asi-Albert (BA) graphs, in which node degree follows a power law \cite{BarAl1999}, generated with the NetworkX  \cite{hag08}  function \texttt{barabasi-albert-graph()} with parameter $m=2$ (each new node is attached to two existing nodes with probability proportional to their existing degree). Graphs with this structure are common in networks of associations, such as social networks or webs of internet sites, and are of particular interest in many graph-structured domains.

In many applications of graph matching, we may be searching for nodes of a given \textit{type} without knowing their detailed \textit{identity}. For example, our data may consist of a detailed graph of financial transactions, in which all participants (businesses, banks, individuals) are fully identified, and our query may be looking for a node of type "person" whose identity is unknown, but who has contacts with other, known individuals and who deals with a specific bank. To support this structure, we furnish each node with an alphabetic \textit{label}, drawn from a fixed vocabulary $\textbf{L}$, representing its type, and a numerical $detail$ that is distinct for each different node of a given type. A node in the query with detail = 0  will match any node of the same type (that is, the same label) in the data. Thus A23 and A42 might be specific, known people, while A0 would be a person whose identity is not known, and who may match either A23 or A42. The labels and details of the kernel are preserved in the query and the data, and other nodes in both query and data are generated with distinct label-detail identifiers. In most of our experiments, $|\textbf{L}| = 100$, and all nodes have non-zero detail. Ablation experiments (Section~\ref{sec:ablation}) set some proportion of the details to 0, and also explore the effect of vocabulary size.

\subsection{Classes and Pseudocode}
In this section, "select from $\textbf{X}$ by $y$" refers to roulette selection from the elements of set $\textbf{X}$ weighted by attribute $y$ of those elements.

ASSIST has three main classes of objects: the \textit{nodes} $\textbf{N}_q$,  $\textbf{N}_d$ of the query and data graphs, their undirected edges $\textbf{E}_q$, $\textbf{E}_d$ edges , and the swarming agents $\textbf{A}$. $\textbf{N} \equiv \textbf{N}_q \cup \textbf{N}_d$, and $\textbf{E} \equiv \textbf{E}_q \cup \textbf{E}_d$.

Each node has the following attributes:
\begin{itemize}
	\item $pherLevel$ (initially 1.0)
	\item $nbrPhers$, an array of length $|\textbf{L}|$ with the total pheromone for nodes with each label among the node’s neighbors
	\item $peers$,  an array of peers in the other graph, each with a $weight$ computed as the cosine distance between the $nbrPhers$ arrays of the peered nodes	
	\item $liveEdges$, number of edges adjacent to the node with \linebreak $pherLevel > 0$ 
\end{itemize}

Each edge has $pherLevel$ (initially 0.0).

Each agent has the following attributes:
\begin{itemize}
	\item $start$, the node on which the agent started its search
	\item $location$, the node where the agent is currently located
	\item $mode$ in $\{1, 2, 3, 4\}$ tracking the agent's progress through the search algorithm outlined below
	\item $history$, a sequential list of nodes and edges traversed in both graphs; $history[0]$ is starting $location$
	\item $startNbr$, the label of the largest element of the starting node’s $nbrPhers$ (selected from $history[0].nbrPhers$ by roulette)
\end{itemize}

ASSIST is implemented in  Repast  \cite{North2013},  which measures time in \textit{ticks}.  Algorithm~\ref{alg:doOneTick} shows the sequence of actions in each tick.

		\begin{algorithm}
		\caption{Sequence of events in each Repast tick}
		\label{alg:doOneTick}
		\begin{algorithmic}[1]
			\Procedure{doOneTick}{}
			
			\ForEach{Node \Id{n} in  $\textbf{N}$}
			\Statep{update $n.liveEdges$}
			\EndFor
			
			\ForEach{Node \Id{n} in  $\textbf{N}_q$}
			\If{ $|n.peers| > 0$}
			\Statep{Update $weight$ of each peer}
			\Statep{Initialize $1 + 2 \cdot liveEdges$ agents with $start \gets history[0] \gets location \gets n$}
			\EndIf
			\EndFor
			
			\ForEach{Agent \Id{a} in  $\textbf{A}$}
			\Statep{Execute procedure \textsc{step}()} [Deposits pheromones]
			\EndFor
			
			\ForEach{Node \Id{n} in  $\textbf{N}$}
			\Statep{Query  neighbors to update  $n.nbrPhers$}
			\EndFor
			
			\ForEach{Edge \Id{n} in  $\textbf{N}$} [Evaporate node pheromones]
			\Statep{$n.pherLevel \gets n.pherLevel \cdot 0.9$}
			\Statep{$n.nbrPhers \gets n.nbrPhers \cdot 0.9$}
			\EndFor
			
			\ForEach{\Id{e} in  $\textbf{E}$}[Evaporate edge pheromones]
			\Statep{$e.pherLevel \gets e.pherLevel \cdot 0.9$}
			\EndFor				
			
			\EndProcedure
		\end{algorithmic}
	\end{algorithm}

The main agent method, invoked in Algorithm~\ref{alg:doOneTick} line 9, is Algorithm~\ref{alg:step}.

	\begin{algorithm}
		\caption{Step method executed by each agent to search for matching edges}
		\label{alg:step}
		\begin{algorithmic}[1]
			%	\State \textbf{method} \textsc{step}()
			\Procedure{step}{}
			\If{$mode = 1$}  [at home in query, seeking peer in data]
			\If{$|location.peers| > 0$} 
			\Statep{select from $peers$ by $weight$}
			\Statep{set $location$ to selected peer}
			\Statep{append peer to $history$}
			\Statep{$mode \gets 2$}
			\Else
			\Statep{deallocate agent}
			\EndIf
			
			\ElsIf{$mode =2 $} [on peer in data, seeking neighbor in data]
			\If{$location$ has nonzero $nbrPhers$ for label $startNbr$}
			\Statep{select from neighbors with this label by $pherLevel$}
			\Statep{$location \gets$  selected neighbor}
			\Statep{append neighbor and traversed edge to $history$}
			\Statep{$mode \gets$ 3}
			\Else
			\Statep{deallocate agent}
			\EndIf
			
			\ElsIf{$mode = 3$} [on neighbor in data, seeking peer in query]
			\If{$|location.peers| > 0$} 
			\Statep{select from $peers$ by $weight$}
			\Statep{set $location$ to selected peer}
			\Statep{append peer to $history$}
			\Statep{$mode \gets$ 4}
			\Else
			\Statep{deallocate agent}
			\EndIf
			
			\ElsIf{$mode = 4$}  [back in query, seeking start node]
			\If{$start$ is neighbor of $location$}% [agent has completed a circuit]
			\Statep{append edge from $location$ to $start$ to $history$}
			\Statep{increment $pherLevel$ on each node and edge in $history$ by 0.1}
			\EndIf
			\Statep{deallocate agent}
			
			\EndIf
			
			\EndProcedure
			
			%\State \textbf{end method}
		\end{algorithmic}
	\end{algorithm}

\subsection{Initialization}
\label{sec:init}

In initialization, ASSIST ingests the query and data graphs, finds nodes common to them both (a process we call "peering"),  and initializes their pheromones.

\textit{Peering} identifies nodes in one graph that match nodes in the other (matching subgraphs of size 1). This process corresponds to the node label filter used by \cite{RN16}, and is motivated by the observation that all the nodes in any shared subgraph must match between the graphs. Matching requires the peered nodes to have the same label, and (unless one of them has detail 0) also the same detail. 

Peering considers each node in the query. We load the data graph as a tree organized according to our data model (Section~\ref{sect:datamodel1}), so data access time is logarithmic,  for overall time complexity $O(q\cdot log(d))$. In practice, peering a 100 node query against a $10^6$ node data graph requires a median time of 38 ms, which exceeds the median match time of 6 ms for this scenario, but is not overwhelming.

Nodes in the query and data without peers are \textit{pruned} (removed from $\textbf{N}$), and edges incident on them  are removed from $\textbf{E}$.

Next,  $pherLevel \gets 1.0$ on all  retained nodes, allowing them to initialize $nbrPhers$. Each time agents update pheromones on visited nodes (Algorithm~\ref{alg:step} line 29, invoked in Algorithm~\ref{alg:doOneTick} line 9), nodes update $nbrPher$s (Algorithm~\ref{alg:doOneTick}, line 11).

A node's $nbrPher$ tells a resident agent which labels it can access from there. The weight of a peering between a query node and a data node is the cosine distance between their $nbrPher$s. The existence of a peering does not change over a run, but its weight does change as pheromones evaporate and accumulate.

\subsection{Matching a Single Edge}

Each agent starts on a peered node in the query and seeks a path corresponding to the dashed loop in Figure~\ref{fig:circuit}. A single circuit requires four ticks, advancing through the four values of \textit{mode}. Line numbers reference Algorithm~\ref{alg:step}.
\begin{enumerate}
	\item It first moves from a node in the query to one of its peers in the data, selected from $peers$ by $weight$ (arrow 1, lines 3-7).
	\item It seeks a neighbor of the peer with label $nbrPher$ (arrow 2, lines 11-15). 
	\item From this node, it seeks a peer back in the query, and if successful, moves to it (arrow 3, lines 19-23). 
	\item Finally, it seeks an edge back to its starting node (arrow 4, lines 27-30).
\end{enumerate}
The moves between the two graphs (arrows 1 and 3) must match both label and detail, while arrow 2 (within the data) needs only match the label, and arrow 4 must arrive back at the starting node.

As it seeks such a circuit, the agent maintains a history of the nodes and edges it has visited. Exploring such a path takes constant time, and can be pursued by many agents in parallel (though our current implementation is serial). An agent that completes all four steps deposits pheromones on the nodes and edges it has visited (Algorithm~\ref{alg:step}, line 29). At the end of each tick, we evaporate the pheromone on all edges and nodes (Algorithm~\ref{alg:doOneTick}, lines 12-16). Thus nodes and edges that participate in successful circuits accumulate pheromone, attracting more agents in subsequent iterations, while the pheromone on others evaporates.

Matched nodes are a subset of peered nodes. Two nodes (one in the query, the other in the data) are \textit{peered} if they describe the same entity, but they are \textit{matched} only if they are peered \textit{and} are  part of a complete circuit (Figure~\ref{fig:circuit}). Similarly, two edges are matched if their endpoints are matched, which implies that they are both part of a successful circuit. 

\begin{figure}[h]
	\centering
	\includegraphics[width=0.6\linewidth]{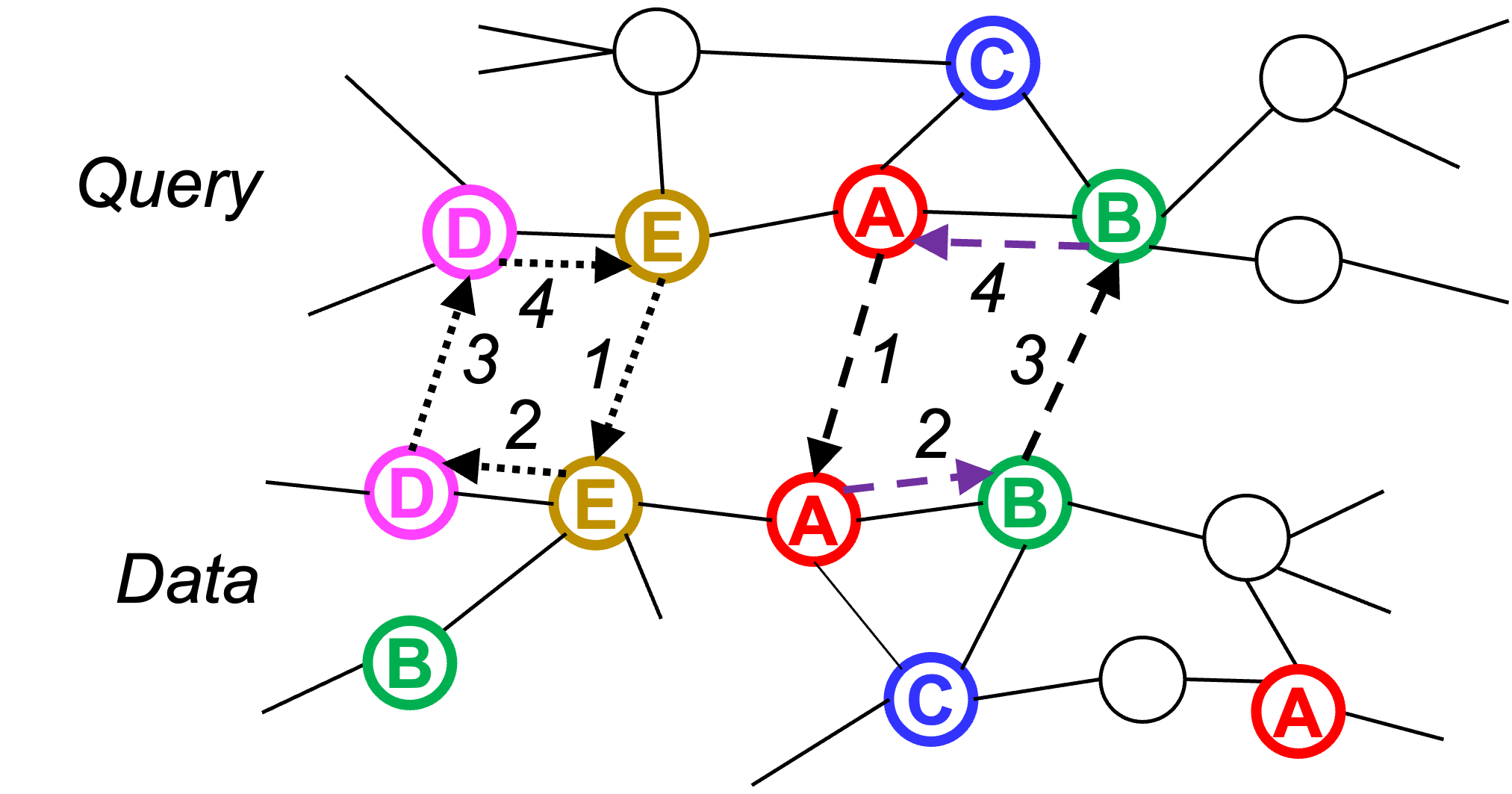}
	\caption{Merging matched edges into larger subgraphs}
	\label{fig:merging}
	%\Description{Merging matched edges into larger subgraphs}
\end{figure}

A single agent circuit identifies only a single shared edge, but many agents stochastically repeating this behavior can merge them into larger shared subgraphs. Figure~\ref{fig:merging} shows another circuit, identifying edge \textit{D-E} as shared. Each of these circuits increases the pheromone strength of its endpoints and their shared edge. As node \textit{E}'s pheromone increases with successive circuits, A's $nbrPher$ augments the value for label \textit{E}, just as successive circuits validating \textit{A} and \textit{B} increase the strength of label \textit{A} in \textit{E}'s $nbrPher$. As a result, the probability increases that agents visiting \textit{A} will consider \textit{E} as well as \textit{B}, and agents visiting \textit{E} will consider \textit{A} as well as \textit{D}, marking the edge \textit{E-A} in each graph as matched and yielding a larger matched subgraph \textit{D-E-A-B}.  A similar process will discover the match between the edges \textit{C-A} and \textit{C-B} in the two graphs.

Repeated visits by many agents reinforce these pheromones, while pheromones on other edges and nodes evaporate, singling out matching nodes and edges. This incremental assembly of smaller subgraphs into larger ones illustrates the power of stigmergy to coordinate the independent efforts of individual agents.

\subsection{Detecting Termination and Retrieving Results}
\label{sec:termination}

ASSIST's matching process is continuous and emergent. For it to be useful in practice, we need to 1) measure its performance on problems of varying complexity, 2) tell when it has converged, and 3) retrieve the matching subgraphs it has found.

We illustrate with a scenario involving a kernel of size 10 embedded in a query of size 30 and a data graph of size 300.

We measure ASSIST's \textit{performance} using the kernel. Because we know the kernel, we can tell when all of its nodes and edges have been identified, by comparing the edges traversed by successful agents in the query with the edges in the kernel. In the example given here, ASSIST discovered the kernel at tick 8. Even on more complex graphs, in general it is discovered in 20 or fewer ticks. For experimental purposes, most of our results report the time required to retrieve the kernel. But this technique does not address the second and third requirements, in cases where we do not know the identity of subgraphs in advance.

\textit{Convergence} can be detected by the plateauing of various observables in the query, including  the total number of matched edges and nodes discovered by successful agents. 

\begin{figure}[ht!]
	\centering
	\includegraphics[width=1.0\linewidth]{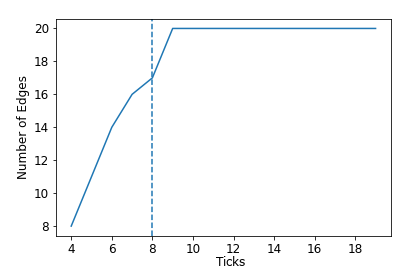}
	\caption{Matched edges by Tick}
	\label{fig:edges}
	%\Description{Matched edges}
\end{figure}

\begin{figure}[h]
	\centering
	\includegraphics[width=1.0\linewidth]{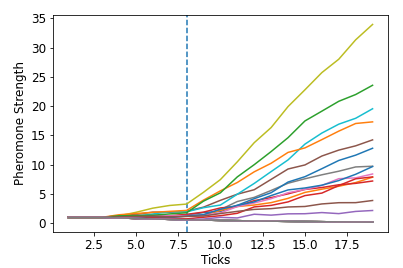}
	\caption{Running average of pheromone levels on query nodes}
	\label{fig:queryra}
	%\Description{Running average of pheromone levels on query nodes}
\end{figure}

Figures~\ref{fig:edges} and \ref{fig:queryra} show the same run, terminated when the number of matched edges has been stable for 20 ticks. The kernel is discovered at tick 8 (the vertical dashed line). The run discovers not only the 10-node, 16-edge kernel, but four more nodes and edges  that are connected with it in the same way in the query and data.

\begin{figure}[h]
	\centering
	\includegraphics[width=1.0\linewidth]{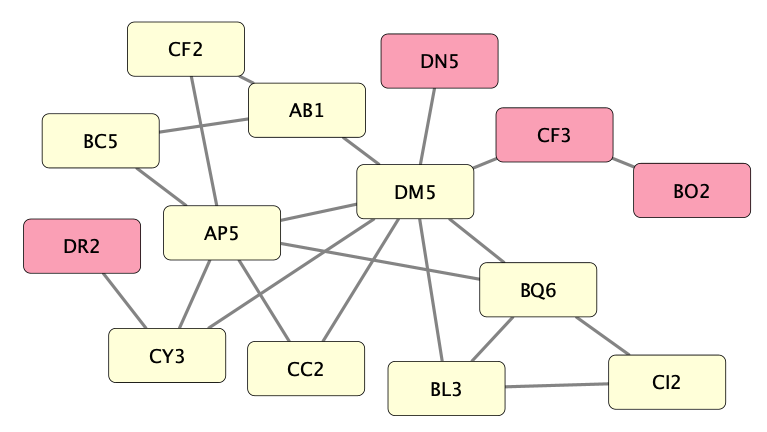}
	\caption{Common subgraph recovered by ASSIST. Yellow: original kernel. Red: other shared nodes and edges.}
	\label{fig:recovered}
	%\Description{Recovered subgraph}
\end{figure}

Figure~\ref{fig:edges} shows the evolution of the number of matched edges. In general, the matched edges may belong to disjoint subgraphs, but in this case, all four are part of the extensions to the kernel shared by the query and the data. Three of the extra edges are detected after the kernel. Figure~\ref{fig:recovered} shows the recovered graph.

The query has 28 peered nodes. Figure~\ref{fig:queryra} shows, for each of these 28 nodes, a running average of the pheromone level on that node. On 14 nodes, the initial pheromone decays away. It accumulates, however, on 14 other nodes, including the ten nodes that form the kernel, clearly separating from the 14 peered but unmatched nodes (all superimposed on the bottom decaying line). In this run, the kernel is discovered soon after matched nodes begin to separate from unmatched ones.

Once the matched edges stabilize, we can \textit{retrieve} the matched subgraph(s), using NetworkX's \texttt{from\_pandas\_edgelist()} function, whose time complexity is linear in the number of discovered edges. For some purposes, it is useful to have a rapid estimate of the largest subgraph discovered at a given point in the process, and we use Cichon's stochastic algorithm \cite{cich11}.

\section{Experimental Results}

\subsection{Dimensions of Interest}

To demonstrate ASSIST, we explore the effect of several variables on the time required to match the kernel ("matching time"). Unless otherwise noted, match times are milliseconds (ms) on a MacBook Pro 18,1 with the Apple M1 Pro chip and 32 GB of RAM, running MacOS 26.1 (Tahoe), measured by calls to \texttt{System.currentTimeMillis()} from \texttt{java.lang.System}. Each run ends when the maximum number of matched edges has not changed for 10 ticks, and we report the time at which the kernel matched (which is often several ticks before matched edges plateaus).

Throughout these experiments, our results are medians over at least five runs with different random seeds, but on the same kernel, query, and data graph for each independent variable. Error bars show the upper and lower quartiles.

The space of interest is huge, and we report only a few results along the following dimensions to illustrate ASSIST's performance.

\begin{itemize}	
	\item Data and Query Size: Given the NP-complete nature of subgraph isomorphism, a primary result of interest is how peering and matching time varies with the size of the query and data graphs. 
	
	\item Common Graph Size: We expect matching time to increase with the size of the largest common subgraph. As noted in Section~\ref{sec:termination}, this may be larger than the kernel. 
	
	\item Query Ambiguity: In our baseline experiments, within each graph, each node's label-detail combination is unique, and matches require matching both label and detail. We expect matching time to increase as we increase the proportion of nodes in the query with detail = 0.
\end{itemize}

\subsection{Data and Query Size}

We expect any algorithm to require more time to process larger query and data graphs. What is the shape of this dependency?

Peering reduces the size of the data graph to the number of peered nodes, which (with unambiguous node identifiers) is bounded by the size of the query, and may be smaller. For example, in one run with $(q,d) = (100, 100)$, the peer set is only 43 nodes. Thus we expect matching time to be independent of data size. 
 
\begin{figure}[h]
	\centering
	\includegraphics[width=1.0\linewidth]{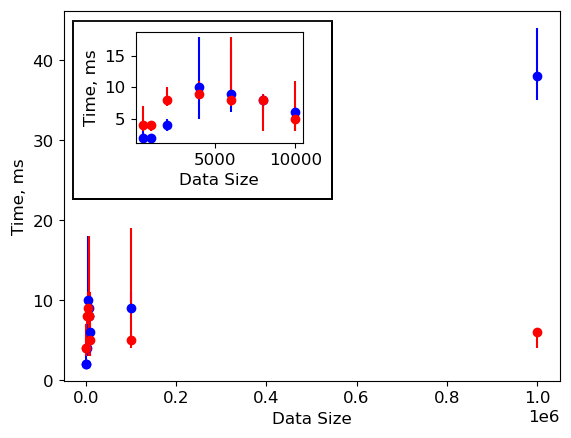}
	\caption{Matching (red) and peering (blue) time for 100 node query by data size}
	\label{fig:matchxdata}
	%\Description{Match time for 100 node query by data size}
\end{figure}

Figure~\ref{fig:matchxdata} shows  matching and peering time for a 100 node query as a function of data size \textit{d}. The inset (for \textit{d} from 100 to 10k nodes) shows that for $d > 3000$, matching time is basically flat, and the full graph shows that this holds for $10^5$ and $10^6$. Under these circumstances, peering dominates matching time for larger data.

\begin{figure}[h]
	\centering
	\includegraphics[width=1.0\linewidth]{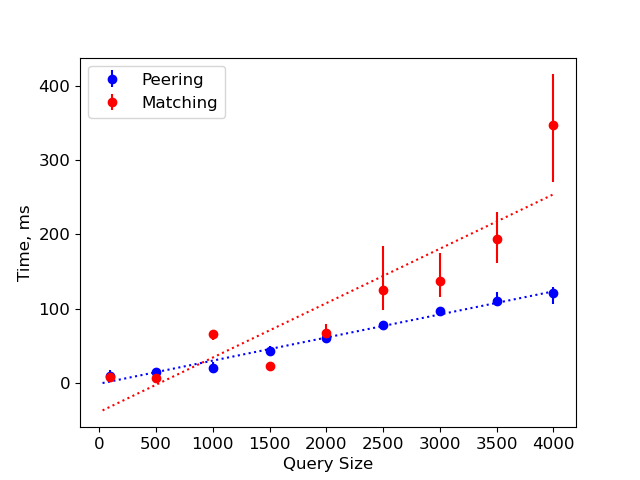}
	\caption{Kernel matching and peering time for 4000 node data by query size}
	\label{fig:matchxquery}
	%\Description{Match time for 4000 node data by query size}
\end{figure}

Figure~\ref{fig:matchxquery} shows kernel matching and peering times for $d = 4000$ as a function of query size \textit{q}.  The dependency in both cases is reasonably linear, though peering grows more slowly than matching.  We expect this dependency. The larger $q$, the larger the set of peered nodes and edges that the agents need to explore.

\begin{figure}[ht!]
	\centering
	\includegraphics[width=1.0\linewidth]{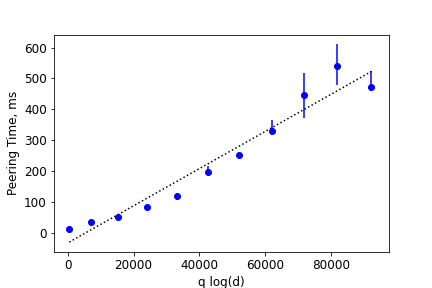}
	\caption{Peering time vs. $q \cdot log(d)$ for $q$ = $d$}
	\label{fig:peervsqlogd}
	%\Description{Match time for 4000 node data by query size}
\end{figure}

We predicted peering time of  $q \cdot log(d)$. Figure~\ref{fig:matchxquery} shows the linear dependence on $q$, but the variation in Figure~\ref{fig:matchxdata} is too high for a useful fit of the $log(d)$ component. Figure~\ref{fig:peervsqlogd} shows that peering time for $q=d \in \{100, 1000, 2000, 3000, 4000, 5000, 6000, 7000, 8000, 9000, 10000\}$ is indeed linear in  $q \cdot log(d)$. 

\subsection{Common Graph Size}

Matching time depends not only on query and data size, but also on the size of the discovered subgraph. The subgraph may be larger than the kernel, and we terminate our runs when the number of matched edges plateaus, in an effort to capture a largest subgraph. We estimate its size with the Cichon heuristic \cite{cich11}. 

\begin{figure}[h]
	\centering
	\includegraphics[width=1.0\linewidth]{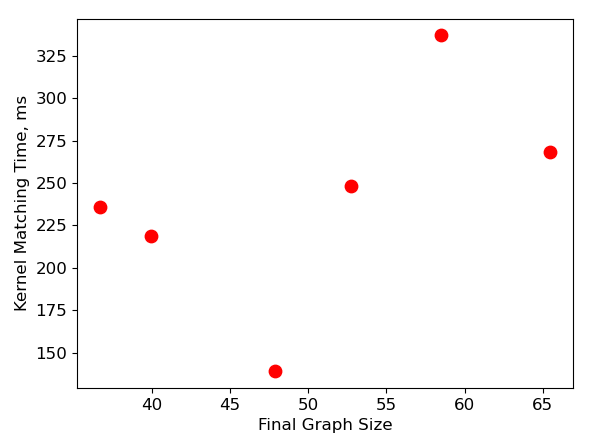}
	\caption{Kernel matching time by largest graph size, 4000 x 4000 scenario}
	\label{fig:fgs}
	%\Description{Match time by largest graph size}
\end{figure}

Figure~\ref{fig:fgs} shows matching time as a function of the overall size of the subgraph discovered at the time the kernel is matched. Each point is a single run, so there are no error bars. The general slope is positive, as expected, but with considerable variation.

\subsection{Query Ambiguity}
\label{sec:ablation}

In many use cases, the query specifies the type of a node, but not its unique identity. Here we explore the effect of ignoring the details on some proportion of the query nodes, with a kernel of size 40, $q = 100, d = 6000$.  When we ignore the detail of a node, we say that we "ablate" it. Ablation requires the algorithm to consider more peers in the data graph for each node in the query than would be needed if we used the detail, and finds subgraphs that satisfy the category (label) of an ablated query node even if the query does not specify the detail.

\begin{figure}[ht!]
	\centering
	\includegraphics[width=1.0\linewidth]{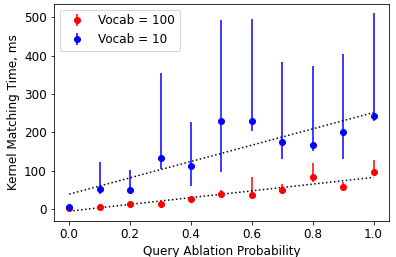}
	\caption{Match time under ablation. Upper curve: vocab = 10. Lower: vocab = 100}
	\label{fig:matchxablation}
	%\Description{Match time by Query Ablation Probability}
\end{figure}

Figure~\ref{fig:matchxablation} shows the impact of varying probabilities of ablation on two scenarios, one with label vocabulary 100 (the same size as other experiments reported here), the other with vocabulary of 10.  In both cases, matching time is linear, though variation is inverse to vocabulary.  Because the labels even without detail carry ten times more information with the larger vocabulary, the peered data graph has fewer nodes and edges than with the smaller vocabulary, and presents an easier search problem. 

ASSIST also supports ablation of the data graph, allowing specific individuals identified in the query to match categories in the data. Space precludes presenting examples here.

\section{Future Work}

These experiments show that ASSIST can find subgraphs in time  $O(q \cdot log(d))$, but leave room for further study. 

\subsection{Parallelization}
The current code is single-threaded. The highly parallel actions of multiple agents hold great promise for multithreading and reimplementation on a GPGPU.

\subsection{Effect of Different Test Graphs}
Evaluation of a stochastic algorithm like ASSIST is appropriately done with random sampling. In the experiments reported here, each scenario consists of a single triple of kernel, query, and data graphs. We run each scenario multiple times with distinct random seeds, varying aspects of the algorithm such as the order in which nodes are explored and the selection of neighbors to explore. But the graphs in the scenario are themselves randomly generated, and it would be worthwhile to expand the sampling, so that for a given kernel, query, and data graph size, multiple different random graphs are explored.

\subsection{Other Random Graph Models}
 Our baseline experiments are with Barab\'asi-Albert (BA) graphs, in which node degree follows a power law \cite{BarAl1999}. Graphs with this structure are common in networks of associations, such as social networks or webs of internet sites, and are of great interest in many potential applications of graph matching. But there are other graph models with different characteristics, including  Erd\H{o}s -R\'enyi graphs \cite{erdos59}  and Watts-Strogatz (small world) graphs \cite{watts98}. These models differ in characteristics such as distribution of node degree, average path length, and clustering coefficient. We plan to explore the performance of ASSIST on these and other models.

\subsection{More Complex Matches}

Figure~\ref{fig:circuit} shows the basic matching mechanism. Some problems have additional complexity. For example:

\begin{enumerate}
	\item Node labels might not match exactly. A data graph might a node "bank," while the query has "financial institution." Maintaining an ontology to allow such abstractions is not difficult, but the simple matching mechanism described above would miss the match.
	\item The graph might be directed. For example, in a graph of financial transactions, the edges indicate transactions, and the movement of money from A to B is not the same as movement from B to A. This directedness imposes a time ordering on the edges, which may be recorded either as clock time or as a partial order over the edges.
	\item We may want to allow matches in which a node or edge might be missing entirely in either the query (due to oversight by the analyst) or the data (due to the vagaries of data collection). 
\end{enumerate}

These complications frustrate many other subgraph algorithms, but straightforward extensions to ASSIST can accommodate them. 

\begin{figure}[h]
	\centering
	\includegraphics[width=1.0\linewidth]{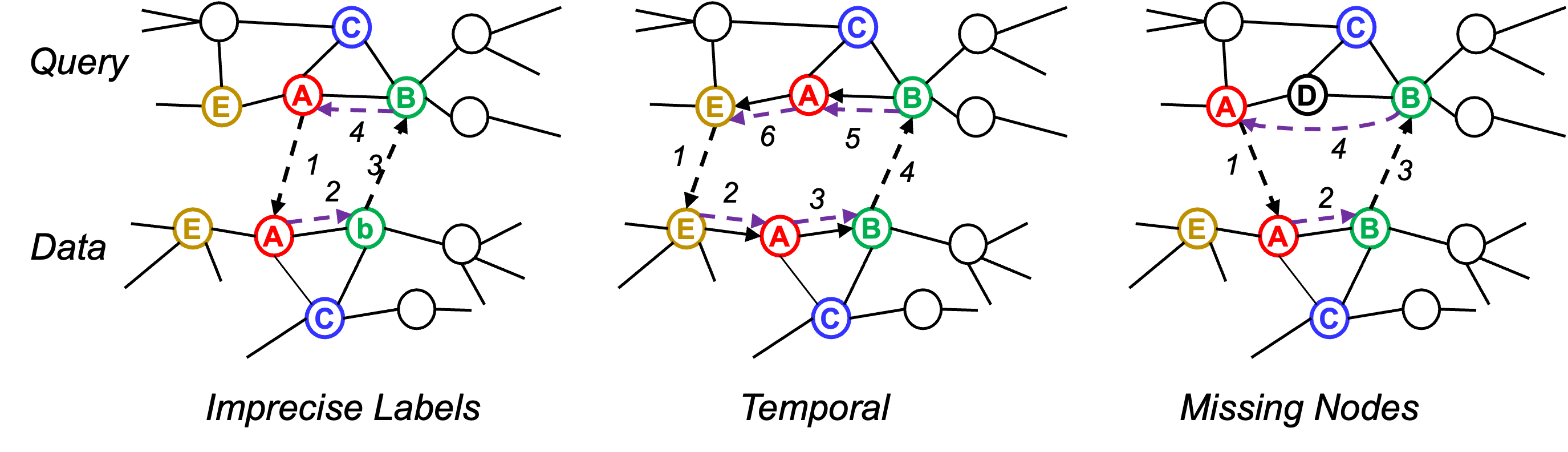}
	\caption{Extensions of ASSIST to more complex matches}
	\label{fig:extensions}
	%\Description{Extensions of ASSIST to more complex matches}
\end{figure}

The query in Figure~\ref{fig:circuit}, repeated in Figure~\ref{fig:extensions} ("Imprecise"), handles case 1 (here, matching ‘B’ with ‘b’) by having agents consult an ontology in case of mismatch to see if one of the nodes subsumes the other. The total pheromone deposited for an imprecise match will be less than that for an exact match.

Figure~\ref{fig:extensions} ("Temporal") handles temporal matches by requiring the agent to traverse two edges in the data graph before returning to the query, remembering the sequence of these edges, and then seeking a sequence of edges in the query with the same order to return home. 

Figure~\ref{fig:extensions} ("Missing") handles missing data by propagating neighbor pheromones across multiple edges, rather than simply sampling adjacent neighbors as in the present implementation. An agent can then sense the presence of an otherwise desirable node not immediately adjacent to its current node and move to it. In this case, as in the case of imprecise data, the pheromone deposited at the end of the circuit will be less than in the case of a perfect match. 

\iftrue
\section{Conclusion}

ASSIST, a swarming stigmergic algorithm, offers an extremely fast heuristic for subgraph isomorphism. After initial peering (which requires time $O(q\cdot log(d))$), matching is linear in query size and constant in data size, much faster than the best previous heuristics, which are quadratic in the number of nodes. In addition, it allows approximate matches, in which a query that specifies only a node's category can retrieve subgraphs that match specific individuals in that category from the data.

In addition to advancing the state of subgraph isomorphism, ASSIST provides a pattern for how stigmergic reasoning can efficiently integrate results produced by multiple agents who are working independently on separate parts of a complex problem.
\fi

%%%%%%%%%%%%%%%%%%%%%%%%%%%%%%%%%%%%%%%%%%%%%%%%%%%%%%%%%%%%%%%%%%%%%%%%

\balance

%%% The next two lines define, first, the bibliography style to be 
%%% applied, and, second, the bibliography file to be used.

\bibliographystyle{ACM-Reference-Format} 
\bibliography{ASSIST}

%%%%%%%%%%%%%%%%%%%%%%%%%%%%%%%%%%%%%%%%%%%%%%%%%%%%%%%%%%%%%%%%%%%%%%%%

\end{document}